\documentclass{aa}  
\usepackage{graphicx}
\usepackage{hyperref}
\usepackage{txfonts}
\usepackage{siunitx}
\usepackage{physics}
\usepackage{orcidlink}

\begin{document} 
    \titlerunning{Separation of gain fluctuations and continuum signals}
    \title{Separation of gain fluctuations and continuum signals in total power spectrometers with application to COMAP}
   \authorrunning{J.~G.~S.~Lunde et al.}
    \author{
        J.~G.~S.~Lunde\inst{1}\fnmsep\thanks{\email{\href{mailto:j.g.s.lunde@astro.uio.no}{j.g.s.lunde@astro.uio.no}}}\orcidlink{0000-0002-7091-8779}
        \and 
        P.~C.~Breysse\inst{2}\orcidlink{0000-0001-8382-5275}
        \and 
        D.~T.~Chung \inst{3}\orcidlink{0000-0003-2618-6504}
        \and
        K.~A.~Cleary\inst{4}\orcidlink{0000-0002-8214-8265}
        \and
        C.~Dickinson\inst{5}\orcidlink{0000-0002-0045-442X}
        \and
        D.~A.~Dunne\inst{4}\orcidlink{0000-0002-5223-8315}
        \and
        J.~O.~Gundersen\inst{6}\orcidlink{0000-0002-7524-4355}
        \and
        S.~E.~Harper\inst{5}\orcidlink{0000-0001-7911-5553}
        \and
        G.~A.~Hoerning\inst{5}\orcidlink{0000-0002-8677-6656}
        \and
        H.~T.~Ihle\inst{1}\orcidlink{0000-0003-3420-7766}
        \and
        J.~W.~Lamb\inst{7}\orcidlink{0000-0002-5959-1285}
        \and 
        T.~J.~Pearson\inst{4}\orcidlink{0000-0001-5213-6231}
        \and 
        T.~J.~Rennie\inst{5,8}\orcidlink{0000-0002-1667-3897}
        \and
        N.-O.~Stutzer\inst{1}\orcidlink{0000-0001-5301-1377}
    }

   \institute{
       Institute of Theoretical Astrophysics, University of Oslo, P.O. Box 1029 Blindern, N-0315 Oslo, Norway
      \and 
       Department of Physics, Southern Methodist University, Dallas, TX 75275, USA
       \and 
       Department of Astronomy, Cornell University, Ithaca, NY 14853, USA
       \and 
       California Institute of Technology, 1200 E. California Blvd., Pasadena, CA 91125, USA
       \and 
       Jodrell Bank Centre for Astrophysics, Alan Turing Building, Department of Physics and Astronomy, School of Natural Sciences, The University of Manchester, Oxford Road, Manchester, M13 9PL, U.K.
       \and 
        Department of Physics, University of Miami, 1320 Campo Sano Avenue, Coral Gables, FL 33146, USA
       \and 
       Owens Valley Radio Observatory, California Institute of Technology, Big Pine, CA 93513, USA
       \and
       Department of Physics and Astronomy, University of British Columbia, Vancouver, BC Canada V6T 1Z1, Canada
   }

    \abstract{
    We describe a time-domain technique for separating $1/f$ gain fluctuations and continuum signal for a total power spectrometer, such as the CO Mapping Array Project (COMAP) Pathfinder instrument. The $1/f$ gain fluctuations of such a system are expected to be common-mode across frequency channels. If the instrument's system temperature is not constant across channels, a continuum signal will exhibit a frequency dependence different from that of common-mode gain fluctuations. Our technique leverages this difference to fit a three-parameter frequency model to each time sample in the time-domain data, separating gain and continuum. We show that this technique can be applied to the COMAP Pathfinder instrument, which exhibits a series of temporally stable resonant noise spikes that effectively act as calibrators, breaking the gain degeneracy with continuum signals.
    Using both simulations and observations of Jupiter, we explore the effect of a $1/f$ prior for the gain model. We show that the model is capable of cleanly separating Jupiter, a bright continuum source, from the gain fluctuations in the scan.
    The technique has two applications to COMAP. For the COMAP observations performing line intensity mapping (LIM), the technique better suppresses atmospheric fluctuations and foregrounds than the COMAP LIM pipeline. For the Galactic COMAP observations, which map Galactic continuum signals, the technique can suppress $1/f$ gain fluctuations while retaining all continuum signals. This is demonstrated by the latest COMAP observations of $\lambda$-Orionis, where our method produces far cleaner maps than a destriper alone, typically reducing the noise power by a factor of 7 on beam scales and up to 15 on larger scales.
    }


   \maketitle

\section{Introduction}

A fundamental challenge in single dish radio astronomy is detecting a faint milli- or micro-kelvin signal on top of a system temperature that is orders of magnitude higher. To accurately measure such faint signals, radiometers employ one or more amplifiers, boosting the faint signal. The stability of the first low noise amplifier (LNA) stage has become a limiting factor in modern radiometer observations \citep[see e.g.,][]{Harper_2018_1f}. Slow fluctuations in the gain provided by the LNA, commonly known as $1/f$ noise, are multiplied by the entire system temperature of the instrument \citep{Weinreb_2014}. As the raw noise performance of these instruments is often at the mK level, a gain stability on the order of $\sim 10^{-5}$ is required for the $1/f$ noise not to dominate the noise budget on long timescales. This temporally correlated noise maps to large scales on the sky, and is often degenerate with the targeted astrophysical signal. This challenge is particularly acute for total power radiometers, which directly measure the product of system gain and total power, offering no inherent mechanism to distinguish a real signal from amplifier gain fluctuations.

Instrument-level mitigations for this problem exist. A common solution is the Dicke switch \citep{Dicke_1946}, in which the receiver quickly alternates between observing the sky and a reference load (e.g., a noise diode) to continuously calibrate the instrument. However, this method is limited by the accuracy of the reference load, which might have its own $1/f$ fluctuations. A traditional Dicke switch also halves the effective on-sky integration time, although this can be mitigated by using a more complex pseudo-correlation radiometer \citep[e.g.,][]{Jarosik_2003, Mennella_2010}. Some suppression of $1/f$ noise can also be carried out in mapmaking, with methods such as destriping \citep{Keihanen_2005_MADAM}, which are most effective for scanning strategies that can quickly cover the entire observed field.

The COMAP Pathfinder \citep{Lamb_2022} is a total power single-dish telescope, observing at $26$--$\SI{34}{GHz}$. The signal from each detector passes through cryogenic low noise amplifiers (LNAs) and ambient-temperature intermediate-frequency (IF) electronics, before digitization into $4096$ frequency channels of $\SI{2}{MHz}$ width.

The Pathfinder system is used for two completely different types of observations. Its primary science is based on CO line intensity mapping (LIM) observations of diffuse emission from the CO(1-0) spectral line at redshifts $z=2.4$--$3.4$ \citep{Lunde_2024, Stutzer_2024, Chung_2024, Dunne2025_COMAP_Stack}. As a secondary science case, the Pathfinder also performs observations of continuum emission from Galactic sources \citep{Rennie_2022,Harper_2025}.

For the CO-LIM observations, $1/f$ gain is already heavily suppressed with a simple linear frequency filter \citep{Foss_2022}. However, this method is not entirely optimal for subtracting continuum contamination, such as foregrounds or the atmosphere. For the Galactic observations, the linear frequency filter cannot be employed, as it would also heavily filter continuum signals. These observations instead relied on a destriper to filter the gain, a method that does not utilize any spectral information about the gain.

In this paper, we propose a method for jointly fitting and separating $1/f$ gain fluctuations and continuum signals. The method relies on an instrument with a system temperature exhibiting frequency-dependent features that are stable over time. The COMAP Pathfinder instrument fulfills this criterion, having a series of resonant noise spikes at specific frequencies. These are stable enough over time to serve as calibrators for the $1/f$ noise, breaking the spectral degeneracy between the gain fluctuations and the continuum signals. Using hourly calibrations of the system noise temperature, we apply this parametric model to COMAP data, fitting it per time sample.

For Galactic continuum science, this model allows separating $1/f$ gain fluctuations from Galactic continuum emission, something that was previously unfeasible. Applied to the CO-LIM observations, which were already capable of filtering $1/f$ gain noise, the model provides a moderate improvement in filtering continuum contamination.

Our method is conceptually similar to the so-called ``pilot-tone'' calibrations, where a strong signal is injected at specific frequencies, which are reserved as calibrators of the $1/f$ noise \citep[e.g.][]{Pollak_2019_gain_stab, Burns_2017}. The technique described in this paper shares a similar fundamental calibration principle but does not rely on injecting a new signal; instead, it utilizes inherent, stable features of the instrument's system temperature.

The paper is structured as follows: Section 2 lays the theoretical foundations of our technique by deriving it from the COMAP data model. Section 3 explores the effectiveness and limitations of the technique through a simple simulation, while Section 4 tests the method on a Jupiter scan. Sections 5 and 6 then demonstrate how the technique can be employed for the CO-LIM and Galactic science cases, respectively. Finally, we conclude in Section 7.

\section{Methodology}
\subsection{The COMAP data model}
\begin{figure}
    \centering
    \includegraphics[width=\linewidth]{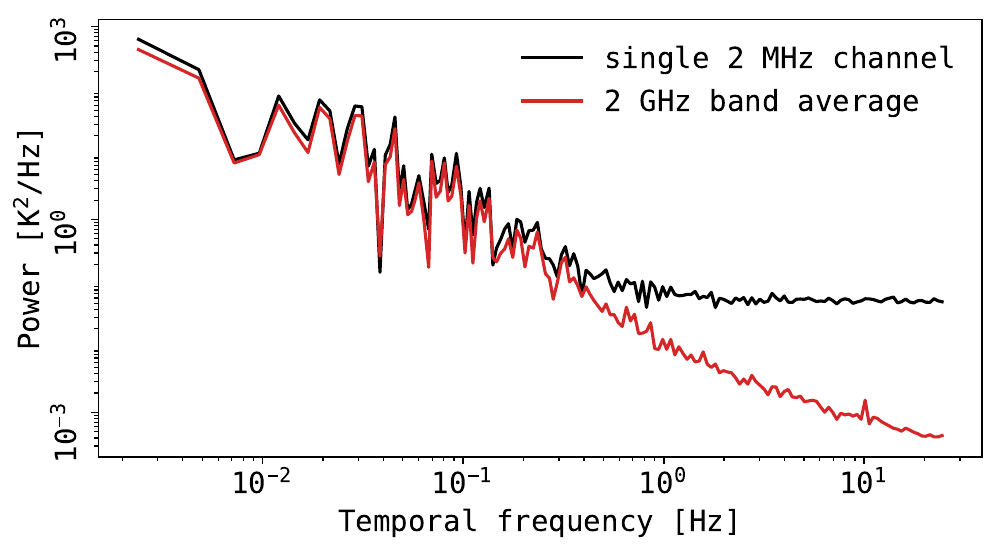}
    \caption{Temporal power spectrum of a single frequency channel (black) and a 1024-channel band-average (red) of a single COMAP scan. The averaging of channels suppresses the spectrally uncorrelated white noise, but not the highly correlated $1/f$ gain noise.}
    \label{fig:sb_vs_single_PS_illustration}
\end{figure}

Throughout the paper, we will use subscripts to denote dependency, meaning that both $d_{\nu,t}$ and $d_{\nu}$ are time-ordered data (TOD) of the same shape, but the former varies with both frequency $\nu$ and time $t$, while the latter is constant in time. All quantities with subscripts can be considered vectors, and there is no special meaning ascribed to capital letters. Variables preceded with a $\delta$ have a time-average of 0.

The digitized signal output of the telescope, $d_{\nu,t}^\mathrm{raw}$, can be written as a product of the system temperature, $T_{\nu,t}$, multiplied by the gain provided by the LNAs, $G_{\nu,t}$, giving the model\footnote{A factor of $k_B \Delta \nu$ is implicitly absorbed by the gain in this model}
\begin{equation}\label{eqn:initial_tod_model}
    d^\mathrm{raw}_{\nu, t} = G_{\nu, t} T_{\nu, t}.
\end{equation}
In this data model, we have assumed that both the gain and the brightness temperature are general, unknown functions of frequency and time. We will now exploit our physical understanding of each of these sources to create a model with fewer degrees of freedom.

\subsubsection{The gain data model}
Ideally, the gain produced by the LNAs would be stable in time such that it is only a function of $\nu$. We know that it is not, and that it fluctuates with a temporal $1/f$ spectrum. This is a significant source of correlated noise for many experiments, including COMAP (see Fig.~\ref{fig:sb_vs_single_PS_illustration}). Because the full analog signal from each detector is passed through the same LNA before being digitized into 4096 channels, the $1/f$ gain fluctuations are expected to manifest as a simple multiplicative factor applied to the overall gain, independent of frequency \citep{Weinreb_2014}. In other words, we can write
\begin{align}\label{eqn:gain_model}
    G_{\nu,t} = \bar{G}_\nu(1 + \delta G_t),
\end{align}
where $\bar{G}_\nu$ is some time-average gain, and $\delta G_t$ are dimensionless and frequency-independent fluctuations around this mean. We have observed that this simple relation holds well for the COMAP instrument \citep{Lunde_2024}.

\subsubsection{The continuum data model}
Similarly to the gain, we decompose the system temperature into the average system temperature, $\bar T_\nu$, and any fluctuations around this mean, $\delta T_{\nu,t}$. As we have not yet made any assumptions about the shape of these fluctuations, this is done without loss of generality, and results in the decomposition
\begin{align}
    T_{\nu, t} = \bar{T}_\nu + \delta T_{\nu,t}.
\end{align}
The mean system temperature\footnote{This quantity is often denoted $T^\mathrm{sys}$ in other COMAP publications.} $\bar{T}_\nu$ is typically around $\SI{40}{K}$ for the COMAP Pathfinder instrument. It is dominated by the physical temperature of the receiver and the brightness temperature of the atmosphere.

We now wish to expand our model to explicitly include continuum signals. For this paper, we consider a continuum source to be anything with a linear frequency dependence over \SI{2}{GHz} frequency range. For reference, the full COMAP band spans \SI{8}{GHz}. Using this definition, we split up the temperature fluctuations $\delta T_{\nu,t}$ into the linear continuum signal and any other fluctuations that do not behave this way. This results in the model
\begin{equation}\label{eqn:cont_model}
    T_{\nu, t} = \underbrace{\bar{T}_{\nu}}_{\text{system temp.}} + \underbrace{\delta T_t(1 + \alpha_t \bar\nu)}_{\text{continuum}} + \underbrace{\delta T_{\nu, t}^\mathrm{noise}}_{\text{noise, CO...}},
\end{equation}
where $\alpha_t$ is the spectral slope of the continuum signal in K/GHz, and $\bar \nu = \nu - \nu_0$ is the mean-centered frequency\footnote{Defining the equation in terms of the mean-centered frequency $\bar \nu$ reduces the degeneracy between $\alpha_t$ and $\delta T_t$. If $\nu$ was used, $\alpha_t$ would define the tilt around $\nu=0$, which is typically far outside the model range.}. The term $\delta T_{\nu, t}^\mathrm{noise}$ captures signals that do not have a linear frequency dependence, such as the system noise, and any astrophysical spectral signal. For the COMAP LIM science, this importantly includes the CO signal itself (which, for this technique, can be considered very weak noise).

\subsubsection{Putting it together}
Inserting both the terms for the gain and the system temperature into the data model in Eq.~\eqref{eqn:initial_tod_model}, we get
\begin{equation}
    d_{\nu, t}^\mathrm{raw} = G_{\nu, t} T_{\nu, t} = \bar{G}_{\nu}(1 + \delta G_t)\bar{T}_\nu\left(1 + \frac{\delta T_t + \delta T_t\alpha_t \bar \nu + \delta T^\mathrm{noise}_{\nu,t}}{\bar T_\nu}\right).
\end{equation}

Between each re-pointing, the COMAP CO-LIM observations are normalized with a high-pass filter with a knee frequency of $\SI{0.01}{Hz}$. Similarly, the COMAP Galactic observations are normalized by the median between each re-pointing. The time between repointings is typically 5--7 minutes for CO-LIM observations and 10-20 minutes for Galactic observations, which usually have larger fields.
These normalizations remove the mean gain and system temperature offsets from each channel. Each channel will now also have the same white noise level\footnote{From the radiometer equation, we know that the white noise level of each frequency channel in the raw data is equal to $\bar G_\nu \frac{\bar T_\nu}{\sqrt{\Delta\nu\tau}}$, where $\Delta\nu$ is the width of each channel, and $\tau$ is the integration time ($\SI{2}{MHz}$ and $\SI{20}{ms}$, respectively, for COMAP). After the normalization, the noise level is now the same across frequency channels, and equal to $\frac{1}{\sqrt{\Delta\nu\tau}}$.}. The signal now also, by definition, has a mean of zero over time. Applying this to our data model, we get

\begin{equation}\label{eqn:model2}
    d_{\nu, t}^\mathrm{norm} = \frac{d_{\nu, t}^\mathrm{raw}}{\bar{G}_\nu \bar{T}_\nu} - 1
    = (1 + \delta G_t)\left(1 + \frac{\delta T_t + \delta T_t\alpha_t \bar \nu + \delta T^\mathrm{noise}_{\nu,t}}{\bar T_\nu}\right) - 1.
\end{equation}

If $\bar{G}_\nu$ and $\bar{T}_\nu$ are calibrated on short enough timescales that we can assume $\delta G_t \ll 1$ and $\delta T_t \ll \bar{T}_\nu$, we can neglect the second-order term, and write out the parentheses of Eq. \eqref{eqn:model2} as
\begin{equation}\label{eqn:final_model}
    d_{\nu, t} = \delta G_t + \frac{\delta T_t}{\bar T_\nu} + \frac{\delta T_t\alpha_t \bar \nu}{\bar T_\nu} + n_{\nu,t},
\end{equation}
where we have also inserted $n_{\nu,t} = \delta T^\mathrm{noise}_{\nu,t}/\bar{T}_\nu$. Under the assumption that $\delta T^\mathrm{noise}_{\nu,t}$ is dominated by the radiometer noise stemming from the system temperature of the instrument, its amplitude will be proportional to $\bar T_\nu$, such that $n_{\nu,t}$ is now white noise with the same standard deviation across all channels.

Equation~\eqref{eqn:final_model} can be solved independently per time-sample, as a standard linear model with white (Gaussian) noise and three unknown parameters. These are (i) the gain fluctuations $\delta G_t$, which now manifest as a constant offset across all frequency channels, (ii) the continuum temperature $\delta T_t$, and (iii) the continuum temperature slope, $\delta T_t\alpha_t$. For completeness, we walk through the maximum likelihood solution of these three parameters in Appendix \ref{app:ML_sol}.

\begin{figure}
    \centering
    \includegraphics[width=\linewidth]{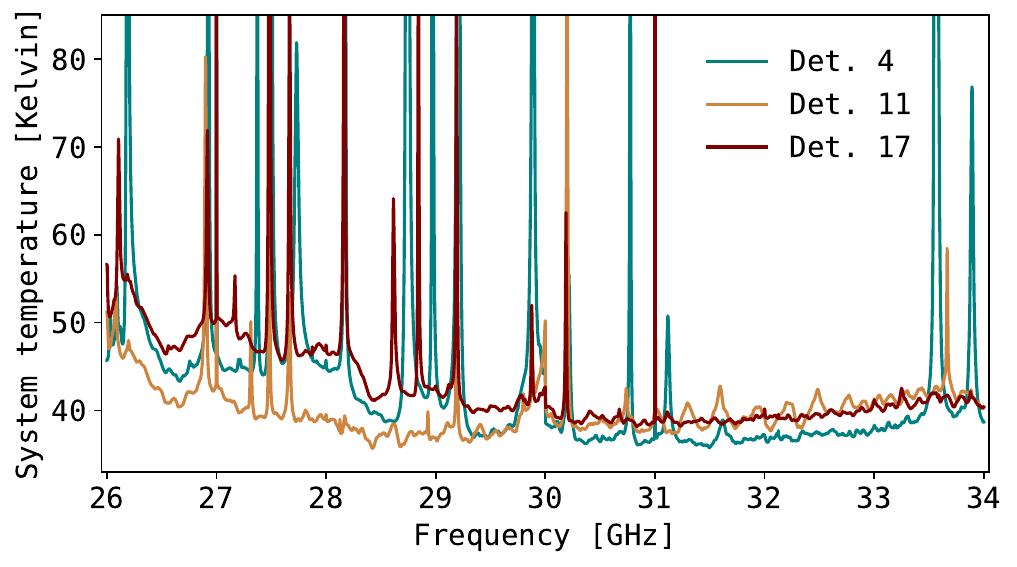}
    \caption{The measured mean system temperature of three detectors for a randomly selected scan. The temperatures deviate somewhat from a flat profile, most notably the sharp spikes, which typically reach brightness temperatures of 100--\SI{200}{K}.}
    \label{fig:Tsys_illustration}
\end{figure}

We will, in practice, be fitting this model to each of the four separate $\SI{2}{GHz}$ sections of the full \SI{8}{GHz} COMAP band. In theory, the model could be fitted across the entire $\SI{8}{GHz}$ range, but in practice, we found that this produced less well-behaved solutions. For the COMAP Pathfinder observations, this equation accounts for virtually all observed power. The only signal sources not accounted for are those with non-continuum frequency dependence, such as the CO signal itself. Features in the atmosphere that are not well approximated as linear on \SI{2}{GHz} scales, like sharp absorption lines, would also not be well-modeled, and result in a residual. The COMAP spectral range of $26-\SI{34}{GHz}$ is generally free of such sharp features in the atmosphere \citep{Paine_2019}. Additionally, any calibration uncertainty, in the form of an incorrectly estimated $\bar T_\nu$, will also result in a residual from actual continuum sources.

\subsection{COMAP system temperature and template degeneracies}
How well we can separate the gain from the continuum depends on the shape of the mean system temperature $\bar T_\nu$. For example, if $\bar T_\nu$ is constant across frequency, $\delta G_t$ and $\delta T_t$ become completely degenerate, and we are left with a two-parameter model of the form
\begin{equation}\label{eqn:model_linear_1}
    d_{\nu,t} = \delta G_t + \alpha_t \bar \nu + n_{\nu,t}.
\end{equation}
This scenario is an advantage or a disadvantage, depending on the use case: For the CO-LIM science, this would have been advantageous, as it would have allowed us to remove both the gain and continuum more easily. For the Galactic science, we want to remove the gain fluctuations, but retain the continuum signal. In this application of the technique, we therefore depend on $\bar{T}_\nu$ not being flat as a function of frequency to separate the two signals.

Figure \ref{fig:Tsys_illustration} shows how $\bar T_\nu$ looks for a selected COMAP scan (and three selected detectors). It is relatively flat across frequencies, except for the spikes. We refer to these as ``system temperature spikes'', and while their exact cause is uncertain \citep{Lamb_2022}, they are actually stable and well-behaved enough (see Appendix \ref{app:Tsys_stability}) that we can utilize them to distinguish the gain and continuum fluctuations.

\begin{figure}
    \centering
    \includegraphics[width=\linewidth]{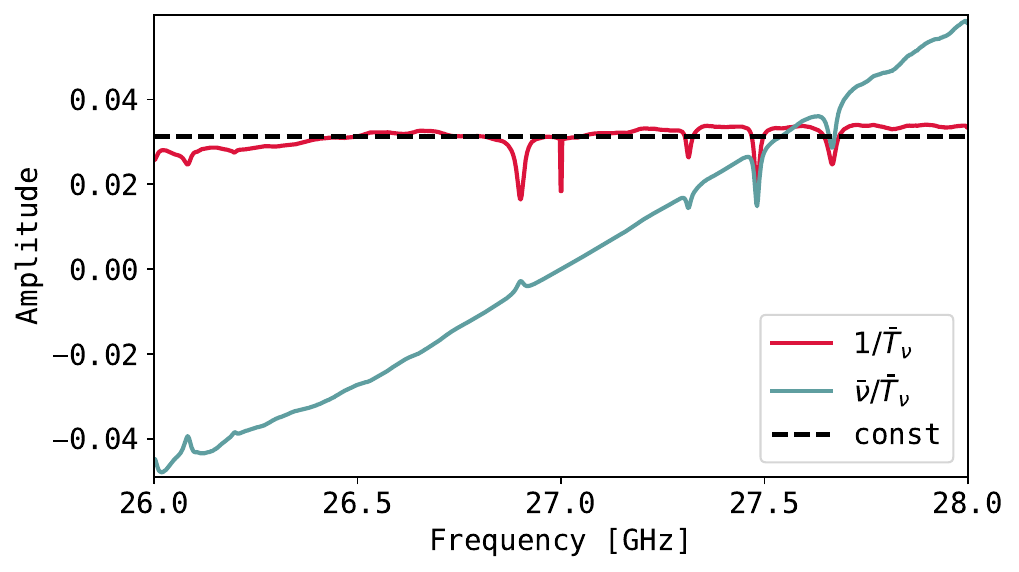}
    \caption{An example of the three frequency templates that are fitted for in our algorithm: (i) The continuum brightness ($1/T\nu$), which is the template for fitting $\delta T_t$. (ii) The continuum slope template ($\bar \nu /\bar T_\nu$), which is the template for fitting $\alpha_t\delta T_t$. (iii) The frequency-constant template, which is the fit for the 1/f gain $\delta G_t$. The flat continuum template ($1/\bar T_\nu$) and the gain template (constant) are highly degenerate.}
    \label{fig:templates_illustration}
\end{figure}
Figure \ref{fig:templates_illustration} shows what the resulting ``template'' functions for the three parameters we fit look like as a function of frequency. The gain template is completely flat, meaning that in our normalized data, we expect a change in gain to manifest as a constant change across all channels. The continuum template, $1/\bar T_\nu$, is also relatively flat, but exhibits downward spikes at the locations of the system temperature spikes. An observed continuum source will therefore look similar to a positive gain fluctuation except in the spiked channels, where the observed power will rise less than it would for a gain fluctuation. Our ability to distinguish $\delta G_t$ and $\delta T_t$, therefore, primarily comes from the behavior of the signal in these spike channels, compared to the flatter parts of the band.

There is a high degree of degeneracy between the gain and continuum templates, with cosine similarities around $0.99$, making them far from orthogonal. Since our data contains noise, this degeneracy in the templates will produce anticorrelated solutions, especially at lower signal-to-noise ratios. The frequency slope of the continuum ($\alpha_t$) does not have this same problem, as it is almost orthogonal to both the other templates. However, because we are fitting the slope to very noisy data over a relatively short band, the signal-to-noise on $\alpha_t$ will be low. Note that the degeneracy between $\delta G_t$ and $\delta T_t$ does not directly imply that either estimate is biased, as the noise causing the degenerate solutions is random. We will further explore this in Sect.~\ref{sec:sims}. However, the degeneracy will lead to excessive noise in the individual parameters, and we will now explore a countermeasure.

\subsection{Prior on the gain}\label{sec:prior}
Because two of the three templates of Eq.~\eqref{eqn:final_model} are not orthogonal, but instead highly degenerate (see Figure \ref{fig:templates_illustration}), the algorithm will often find large and nonphysical solutions to $\delta T_t$, and compensate by letting $\delta G$ have a large amplitude of the opposite sign. To what extent this is actually a problem depends on the use case:
\begin{itemize}
    \item For the CO-LIM observations, we only care about what is contained in the joint solution $\delta G_t + (\delta T_t + \delta T_t\alpha_t \nu)/\bar T_\nu$, as we want to filter both contributions. It is therefore irrelevant if the individual terms are nonphysical on their own.
    \item For the Galactic continuum observations, we want to separate $\delta G_t$ and $(\delta T_t + \delta T_t\alpha_t \nu)/\bar T_\nu$. While noise-induced leakage between the two will not produce a systematic bias (because it is random), it will increase noise in both individual parameters.
\end{itemize}

The solution is to use our insight into the expected temporal power spectrum of the gain fluctuations $\delta G_t$ to place an informed prior on $\delta G_t$. As previously discussed, gain fluctuations arise from well-understood $1/f$ amplifier noise. These have a very well-defined temporal power spectrum of the shape
\begin{equation}
    P(f) = \sigma_0^2 \qty(\frac{f}{f_\mathrm{knee}})^\beta,
\end{equation}
where $\sigma_0$ is the white noise level in the data, $f_\mathrm{knee}$ is the frequency below which the correlated noise dominates, and $\beta$ is the slope of the power spectrum. The specifics of how this prior is implemented into the maximum likelihood solution are outlined in Appendix~\ref{app:ML_sol}. In the equation above, $\sigma_0$ and $f_\mathrm{knee}$ are completely degenerate. In practice, $\sigma_0$ is not a free parameter, but the white noise uncertainty of the data, which can be directly derived either from the data itself or from the radiometer equation.

The two free parameters $f_\mathrm{knee}$ and $\beta$ are typically stable over long periods of time. They can therefore be estimated individually for each detector in the COMAP Pathfinder instrument by fitting the above equation to the band-averaged noise spectra of many individual scans and averaging the results. The resulting best-fit parameters range from $-1.1$ to $-0.9$ for $\beta$, and from \SI{20}{Hz} to $\SI{30}{Hz}$ for $f_\mathrm{knee}$, varying by detector. Note that the $f_\mathrm{knee}$ value is here evaluated for an averaged $\SI{2}{GHz}$ band, as can be seen in Fig.~\ref{fig:sb_vs_single_PS_illustration}, which is why the value is so high. Evaluated compared to the $\sigma_0$ of a single $\SI{2}{MHz}$ band, the $f_\mathrm{knee}$ lies between $\SI{0.6}{Hz}$ and $\SI{1.1}{Hz}$, which can also matches Fig.~\ref{fig:sb_vs_single_PS_illustration}.

The $1/f$ shape of the prior suppresses the small-scale fluctuations in $\delta G_t$. The anti-correlation between $\delta G_t$ and $\delta T_t$ arose because their templates are so correlated that their difference is almost zero, allowing a linear combination of them to be near-zero, which in turn can overfit random noise fluctuations. The prior stops this behavior by disallowing $\delta G_t$ from exhibiting small-scale fluctuations unless doing so would substantially improve the joint fit. Another way of looking at this is that the prior forces $\delta G_t$ to be evaluated not sample-by-sample, but over longer timescales, which essentially improves the signal-to-noise and our ability to distinguish the two templates.

In theory, one could employ a similar prior on the $\delta T_t$ term. A prior on $\delta G_t$ alone is typically sufficient to make both parameters physically meaningful, as we will show in the following sections. We would also need knowledge of the expected power spectrum of $\delta T_t$. For any continuum observation, such as in Galactic science, this poses no clear advantage, and we would not wish to place a prior on the signal we are trying to measure. For the CO-LIM science, the primary target of $\delta T_t$ is the atmosphere, which typically has a predictable power spectrum. The advantage of imposing such a prior is that there would be less loss of large-scale CO signal in the line-of-sight dimension.

\subsection{Comparison to simpler filtering technique}\label{sec:linear_model}
The COMAP CO-LIM pipeline already includes a filtering step that attempts to address continuum and gain contamination in the data \citep{Lunde_2024}, which is simpler than the one we present here. For every time sample, it performs a linear fit across frequencies to the data, assuming that the data takes the form
\begin{equation}\label{eqn:model_linear}
    d_{\nu, t} = a_t \nu + b_t,
\end{equation}
where $a_t$ and $b_t$ are fitted for each time sample. This model is identical to the one we outlined in Eq.~\ref{eqn:model_linear_1}, which was obtained by assuming that the system temperature $\bar T_\nu$ is constant across the considered band. In other words, Eq.~\ref{eqn:model_linear} fails to account for the structure of the system temperature, and this will lead to a continuum residual. \citet{Lunde_2024} partially mitigates this by masking the system temperature spikes before the linear fitting. The gain fluctuations are captured perfectly by the linear model, as it has no dependence on $\bar T_\nu$. Note that when our proposed three-parameter fit is employed, the system temperature spikes must not be masked before the fit, as most of the method's ability to distinguish the gain from the continuum lies precisely in these spikes.

The simpler linear fit has so far been found adequate for the CO-LIM observations, and no meaningful differences were observed in power-spectrum results or null tests at the current sensitivity levels. This is likely partially because the gain fluctuations are, by far, the most important noise source in the Pathfinder observations, and partially because the linear model does an acceptable job at fitting continuum sources when the spike-masking is applied.

\section{Testing with simulations}\label{sec:sims}
\begin{figure}
    \centering
    \includegraphics[width=\linewidth]{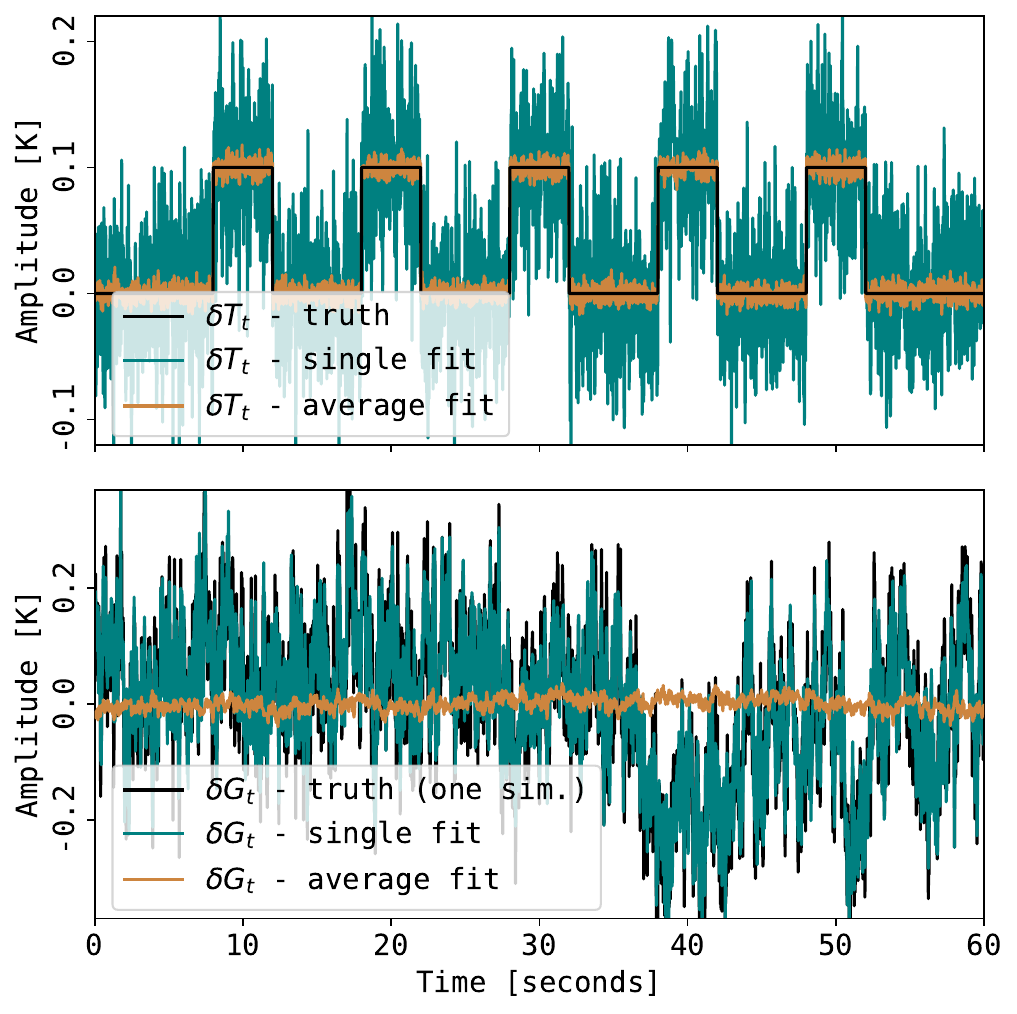}
    \caption{Ground truths and results from the simulations with a correct prior. Top: Ground truth (black) of the top-hat continuum signal, together with a single fit (teal) and the average of 100 simulations (orange), showing that the average appears to approach the ground truth, indicating an unbiased solution. Bottom: The same simulation results for the gain, where the single ground truth (black) is shown together with a single solution (teal). The average of 100 simulations (orange) appears to average towards zero, indicating that there was no leakage of continuum into the gain solution.}
    \label{fig:sims}
\end{figure}

\subsection{Setup}
The purpose of this section is to build intuition for our model by using a one-dimensional simulation in which the true solution for all parameters is known. The simulation assumes that the telescope is performing a one-dimensional scan across the sky and passes over five sharp continuum features, implemented as top-hat functions. On top of this, we add gain fluctuations and white noise, with parameters $\sigma_0 = \SI{0.2}{K}$, $f_\mathrm{knee} = \SI{1.0}{Hz}$, and $\beta = -1.0$, producing noise properties similar to COMAP data. We will refer to the quantities used in the simulations as $\delta G_t^\mathrm{true}$ and $\delta T_t^\mathrm{true}$, while referring to our best-fit estimates of these as $\delta G_t$ and $\delta T_t$. Equation~\eqref{eqn:model2} is used as the data model for the simulations. The simulation uses 1024 frequency channels, 3000 time-samples, and a real $T_{\nu}$ measurement, similar to what is shown in Fig.~\ref{fig:Tsys_illustration}.

We employ three different versions of our algorithm on this simulated data:
\begin{itemize}
    \item No prior on $\delta G_t$. Note that this is also equivalent to an infinitely lenient prior, where $f_\mathrm{knee} = \infty$
    \item A correct prior on $\delta G_t$, where $f_\mathrm{knee}$ and $\beta$ are both set to the same value as $\delta G_t^\mathrm{true}$ was generated from.
    \item A too strict prior on $\delta G_t$, where $f_\mathrm{knee}$ is set to a value 10 times lower than the correct value.
\end{itemize}
The first and last versions allow us to test the effect of a prior that severely misrepresents the true gain fluctuations. 

\subsection{Simulation results}
\begin{figure*}
    \centering
    \includegraphics[width=0.9\linewidth]{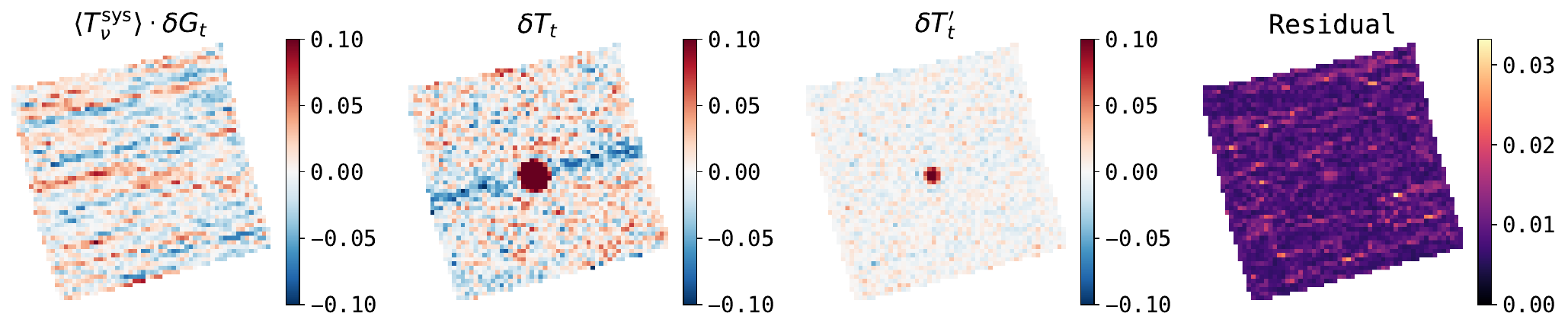}
    \includegraphics[width=0.9\linewidth]{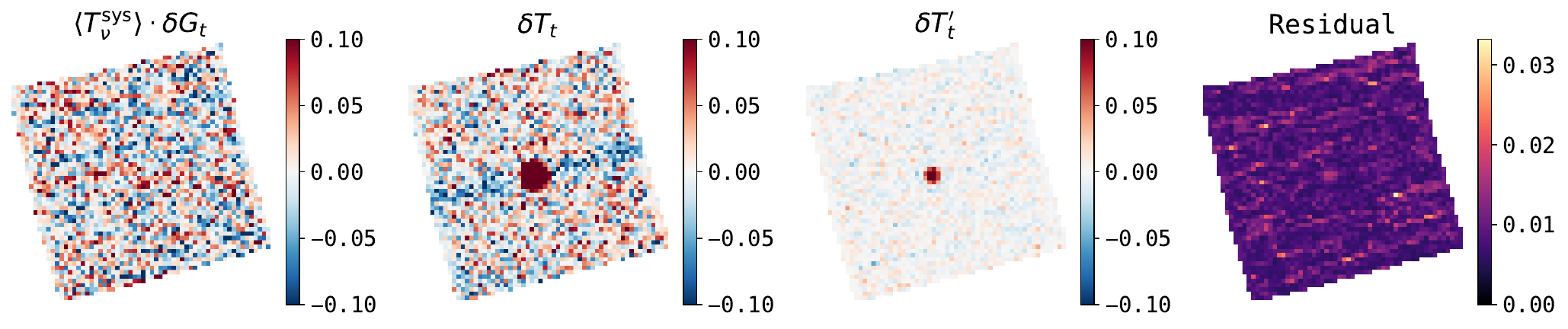}
    \caption{Binned maps of the joint gain-continuum solution to a scan of Jupiter, both with a prior on $\delta G_t$ (top row) and without (bottom row). All maps are in units of $\si{K}$. The first three columns show the binned maps of the three parameters (the gain $\delta G_t$, the continuum brightness temperature $\delta T_t$, and the continuum slope $\delta T_t \alpha_t$), while the last column shows the frequency-averaged absolute residual after subtracting the joint model from the data. The central $\delta T_t$ pixel has an amplitude of $\approx \SI{2.0}{K}$ in both results.}
    \label{fig:jupiter_maps}
\end{figure*}
The top panel of Fig.~\ref{fig:sims} illustrates the input and results of $\delta T_t$ for the simulation configuration with the correct prior. The algorithm correctly recovers $\delta T_t$ without bias. The equivalent solution to $\delta G_t$ is shown in the bottom panel. The fact that the simulations average to zero shows that there is no systematic leakage of $\delta T^\mathrm{true}_t$ into the solution for $\delta G_t$.

\begingroup
\setlength{\tabcolsep}{6pt} % Default value: 6pt
\renewcommand{\arraystretch}{1.2} % Default value: 1
\begin{table}[]
    \centering
    \begin{tabular}{l r r r}
        \textbf{Correct} $\mathbf{\bar T_\nu}$ & No prior & Correct prior & Too strict prior \\
        \hline
        Recovered ampl. & 100\% & 100\% & 100\% \\
        Corr($\delta G_t$, $\delta T^\mathrm{true}_t$) & $\sim 0\%$ & $\sim 0\%$ & $\sim 0\%$\\
        Corr($\delta G_t$, $\delta T_t$) & $-30\%$ & $0.6\%$ & $30\%$ \\
        Std($\delta T_t - \delta T_t^\mathrm{true})$ & $\SI{58}{mK}$ & $\SI{45}{mK}$ & $\SI{94}{mK}$ \vspace{0.1cm}\\
        \textbf{Incorrect} $\mathbf{\bar T_\nu}$ & & & \\
        \hline
        Recovered ampl. & 88\% & 88\% & 92\% \\
        Corr($\delta G_t$, $\delta T^\mathrm{true}_t$) & $4\%$ & $5\%$ & $5\%$ \\
    \end{tabular}
    \vspace{0.1cm}
    \caption{Table showing summary statistics of the simulation, with columns showing different prior strengths. \textit{First row:} Recovered amplitude of continuum sources, relative to true values. \textit{Second row:} Correlation between the gain solution and the true continuum, showing no leakage between them. The first two rows both demonstrate that the recovered solution is unbiased. \textit{Third row:} Correlation between the predicted gain and continuum, demonstrating that only the correct prior creates an uncorrelated solution. \textit{Fourth row:} Standard deviation of predicted continuum with the true solution subtracted, demonstrating that the correct prior gives the least noisy solution. \textit{Bottom two rows:} Simulations repeated with incorrect $\bar T_\nu$, both rows demonstrating that the solution is now slightly biased, with predicted continuum solutions dimmer than the truth.
    }
    \label{tab:sim_table}
\end{table}
\endgroup
The top four rows of table~\ref{tab:sim_table} summarize the solutions to all three simulation versions. For all three configurations, both the fact that the continuum amplitudes were correctly recovered, as well as the lack of correlation between $\delta T^\mathrm{true}_t$ and $\delta G_t$, prove that the continuum solution is unbiased, with no systematic leakage between continuum and gain. This holds even when the incorrect prior is used. The correct prior yields the best-behaved solutions for $\delta G_t$ and $\delta T_t$ in terms of their noise properties. Both without a prior and with a too strict prior, the noise in $\delta T_t$ (after subtracting the true solution) is higher. The results of the third row support these findings, showing that the solutions to $\delta G_t$ and $\delta T_t$ are least correlated when the correct prior is used. When there is no prior on $\delta G_t$, they become highly anti-correlated, inducing excess noise in both solutions. When the prior is too strict, the two solutions become correlated. This happens because $\delta T_t$, which is allowed to vary freely, attempts to model parts of the gain fluctuations, as $\delta G_t$ is too suppressed by the before fully capture these fluctuations alone.

The table does not contain a summary of the properties of the slope $\alpha_t$. The complete continuum slope solution $\delta T_t \alpha_t \bar \nu$ generally exhibits the same qualitative behavior as $\delta T_t$, and results in the same overall conclusions to the points discussed above.

\subsection{Simulations with incorrect system temperature}
The simulations in the previous sections assumed perfect knowledge of the mean system temperature. In practice, $\bar{T}_\nu$ will both slowly change over time and will never be measured to perfect precision. The simulations were therefore repeated, but using a $\bar{T}_\nu$ for creating the simulations that is different from the $\bar T_\nu^\mathrm{approx}$ used when fitting the three-parameter model.

We introduce three simultaneous errors in the incorrect $\bar{T}_\nu^\mathrm{approx}$, all meant to mimic slightly exaggerated realistic failures: (i) We use a real COMAP calibration, just as with the correct $\bar{T}_\nu$, but from the previous calibration, 1 hour before. (ii) We add random white noise to $\bar{T}_\nu^\mathrm{approx}$, with a $\sigma$ about $1\%$ of the amplitude of $\bar{T}_\nu$. (iii) We add a sine wave of about $2\%$ of the amplitude of $\bar{T}_\nu$, to emulate a standing wave between the calibration vane and the detectors, something we observe in real calibrations. The simulations and the three different prior scenarios are then repeated.

The bottom two rows of table~\ref{tab:sim_table} summarize the results of these simulations. The amplitude of the continuum sources is no longer correctly recovered, being biased by $~\sim 10\%$. This is explained by the fact that the best-fit gain solution $\delta G_t$ and ``true'' simulated continuum $\delta T^\mathrm{true}_t$ are now slightly correlated, showing that some continuum $\delta T_t$ now leaks into the gain solution $\delta G_t$. This indicates that the solutions are quite sensitive to an incorrectly estimated mean system temperature. For reference, in Appendix~\ref{app:Tsys_all_uncertainties} we show that the expected measurement uncertainties on $\bar T_\nu$ are around $0.1\%$, while the relative area under the spikes (which is the most crucial quantity for our constraining power) changes on average by about $2.1\%$ between hourly calibrations.

The errors to $\bar T_\nu$ should be divided into two types. If $\bar T_\nu$ retains its overall shape, but is just generally over- or under-estimated, the continuum solution will also be over- or under-estimated, respectively. This is the expected type of error, e.g., from drifts in the instrument's overall system temperature on faster timescales than our calibration can capture. While it will produce a bias in individual scans, it will even out over time.

The second potential error in $\bar T_\nu$ is the addition of an incorrect structure, as in the noise and standing waves we added in the simulations. This will almost always lead to an \textit{underprediction} of the continuum signal (unless the added term has a very specific correlation with the system temperature spikes). The intuitive explanation for this underprediction is simply that the resulting distorted template will tend to be less correlated with what the real continuum signal looks like than the correct template. The relative bias induced by these errors will not depend on the continuum signal strength, as they arise from a template mismatch rather than a lack of signal-to-noise.

\section{Testing on Jupiter scans}
Jupiter is a bright point source of continuum emission and is regularly observed by the COMAP Pathfinder, making it an ideal test case for our algorithm. We employ the algorithm on a single 5-minute scan of Jupiter. This is done both with and without a prior, and using the per-detector prior parameters ($f_\mathrm{knee}$ and $\beta$) described in Sect.~\ref{sec:prior}.

Figure \ref{fig:jupiter_maps} shows the results of the Jupiter analysis. First, we see that the algorithm is very successful at isolating Jupiter in the $\delta T_t$ and $\delta T_t \alpha_t \bar \nu$ terms, with no apparent leakage into $\delta G_t$. This is also evident from the residual, which shows barely any visible footprint of Jupiter. The amount of leakage into the residual, measured as the ratio of the residual map to the continuum map, peaks at $<1.0\%$ in the central pixels. This also indicates that the system temperature of COMAP is measured quite accurately. In Sect.~\ref{sec:sims} we ran simulations with an incorrect $\bar T_\nu$, and found a $10\%$ bias in the continuum solutions, far larger than what the Jupiter results show.

It is also evident from the figure how the prior on $\delta G_t$ helps immensely in correctly separating the power between $\delta G_t$ and $\delta T_t$. We note again that the joint solution $\delta G_t + \frac{\delta T_t}{\bar T_\nu} + \frac{\delta T_t\alpha_t \bar \nu}{\bar T_\nu}$ is barely impacted by the prior, as can be seen from the fact that the residual is virtually identical with both setups. However, without the prior, the individual solutions to $\delta G_t$ and $\delta T_t$ become much noisier, and therefore less useful on their own.

Looking at this solution from the perspective of the Galactic science, where the continuum is the signal of interest while we want to filter away the gain, this result illustrates how the algorithm is capable of capturing a substantial amount of correlated noise into $\delta G_t$. At the same time, there is virtually no leakage of Jupiter into the gain term. This demonstrates its capability as a 1/f gain noise removal tool. Still, it should also be noted that Jupiter is an exceptionally bright continuum signal, providing the algorithm with ample signal to disentangle the quite degenerate template parameters.

\section{Applications to LIM}

For the CO-LIM science case, the purpose of this algorithm is to remove both gain and continuum fluctuations from the data. This presents an improvement to the current COMAP CO-LIM pipeline approach, described by Eq.~\eqref{eqn:model_linear}, which does not fully capture continuum fluctuations. Even so, applications of this technique to the COMAP CO-LIM dataset have shown no conclusive improvement to null tests or final power spectrum results, and the COMAP Season 2 results therefore employed the conceptually more straightforward model outlined in Section \ref{sec:linear_model}. 

The lack of improvement is likely due to continuum emission not being a significant source of systematic error or residual correlated noise beyond what the simpler linear filter can handle. However, several of these continuum sources are not expected to integrate down as noise. This includes astrophysical foregrounds such as the CMB, but also sidelobe ground pickup, as the fields take the same path across the sky every day, potentially picking up a sidelobe signal that maps onto the COMAP field coordinates. As the sensitivity of COMAP increases, both with an increased number of raw observational hours and increased data retention, these continuum sources could start rising above the noise floor. At this point, the suboptimal linear filter model might become problematic. Even though this is uncertain, it means that our technique might become necessary for the COMAP CO-LIM pipeline in the future. This section will therefore further explore the effectiveness of this technique as a continuum filtering tool.

Our method could also be tuned to minimize the loss of large-scale modes in the astrophysical CO signal, as discussed in Sect.~\ref{sec:prior}. The primary target of the continuum term when used for CO-LIM observations is the atmosphere. Atmospheric fluctuations, like gain fluctuations, exhibit a $1/f$-like temporal power spectrum with an even steeper slope. A prior could therefore be placed on the $\delta T_t$ term as well, not for the purpose of reducing the degeneracy with $\delta G_t$, but to reduce the large-scale signal loss from the filtering. We have not pursued this specific application in the present work.

\begin{figure}
    \centering
    \includegraphics[width=\linewidth]{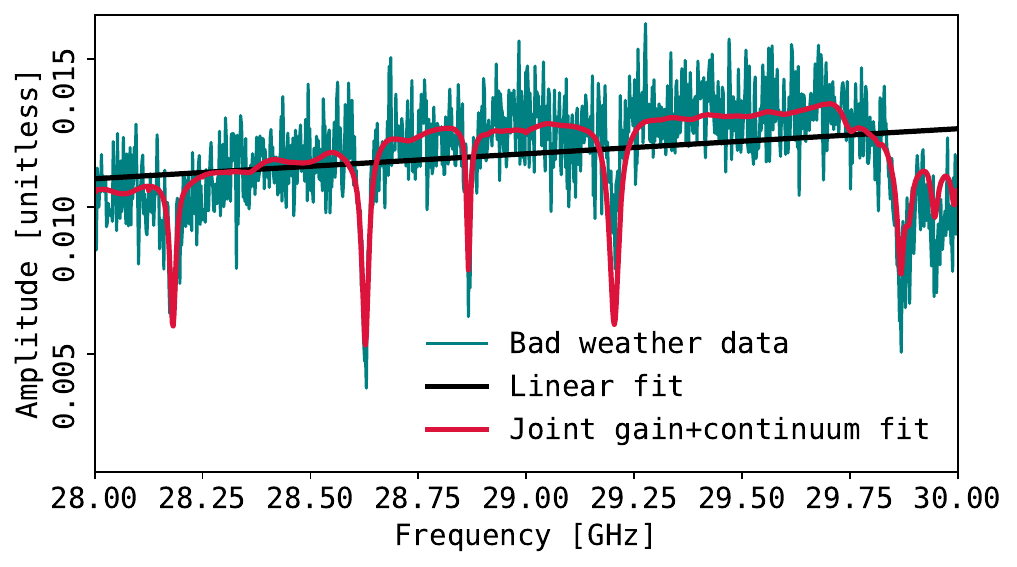}
    \caption{Figure showing the normalized amplitude of data taken during bad weather conditions (teal), which should exhibit strong changes in continuum emission, and therefore be well-modeled by our method. On top is the naive linear fit to the data (black) and the joint fit described by this paper (red). All three lines are averages over 1 second of observation.}
    \label{fig:bad_weather}
\end{figure}
\subsection{Bad weather scan}
Similar to looking at Jupiter, another way of observing a strong continuum source is to look at a scan performed during bad weather conditions, as clouds are expected to have continuum-like emission. Note that these types of scans are not normally included in the COMAP CO-LIM analysis anyway \citep{Foss_2022, Lunde_2024}, and this serves primarily as an illustration of the algorithm under extreme conditions. The method could, however, push the boundary for acceptable weather conditions.

\begin{figure*}
    \centering
    \includegraphics[width=0.9\linewidth]{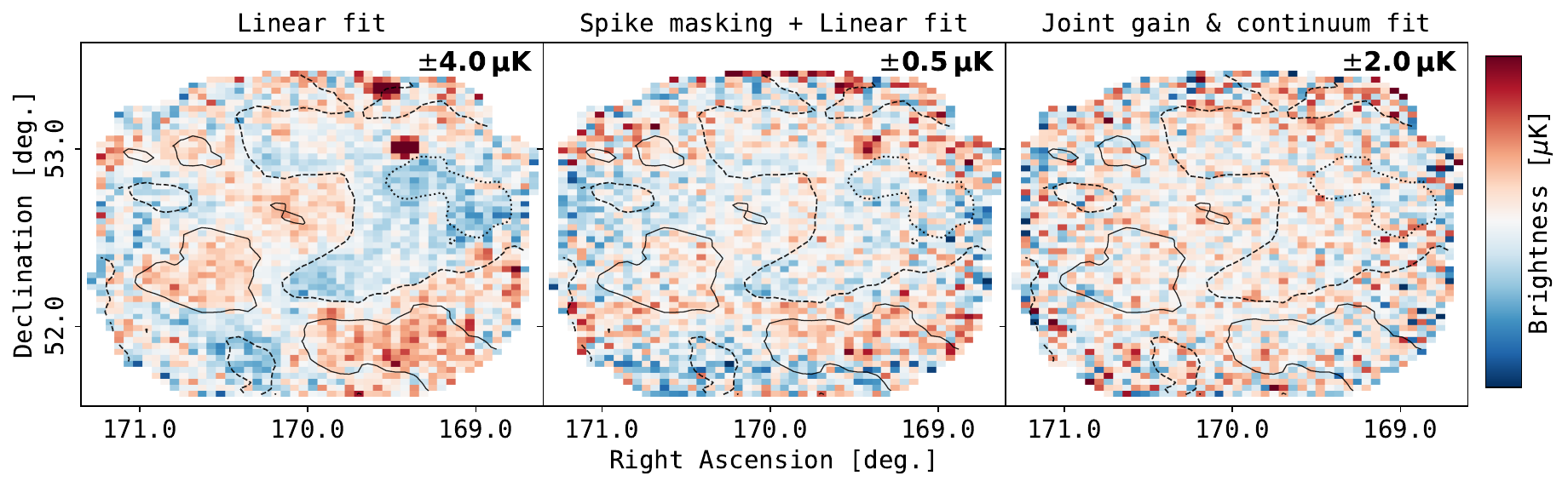}
    \caption{Figure showing COMAP Field 2, averaged over both detectors and all frequency channels. Dotted, striped, and solid contour lines, respectively, show -1.25, 0, and +\SI{1.25}{mK} thresholds of CMB anisotropies from \textit{Planck}. When the simple linear fit is used (left plot), a correlation between the inverse-variance and continuum residual causes the CMB, as well as two point sources, to leak through. Tightly masking excess $\bar T_\nu$ values before applying the linear filter (middle plot) considerably reduces the residual. The joint fit technique described in this paper (right plot) has no visible residual without any masking of $\bar T_\nu$ values. The individual colorbar limits are shown in the top right of each panel. The correlations between the three maps and the CMB template are 41.5\%, 9.7\%, and 0.8\%, respectively, after masking the point sources.}
    \label{fig:CMB}
\end{figure*}

Figure \ref{fig:bad_weather} shows the result of applying the joint-fit algorithm on a scan with bad weather. The result has been averaged over 50 time samples (1 second) to reduce noise. As a reference, the figure also shows the simpler linear fit described by Eq.~\eqref{eqn:model_linear}, and it is clear that the joint gain-continuum fit is a much better model for the data, while the linear fit fails to capture the shape of the continuum signal.

\begin{figure*}
    \centering
    \includegraphics[width=\linewidth]{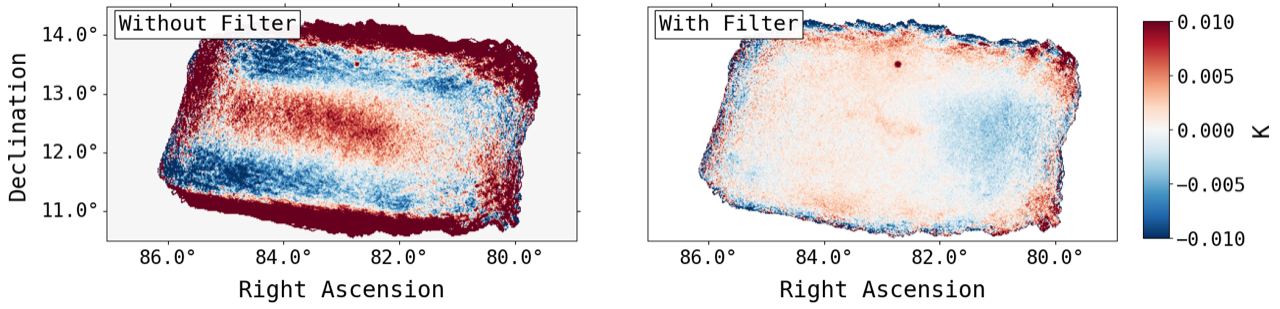}
    \caption{Maps showing the LDN 1582 filament in the $\lambda$-Orionis region presented in \citet{Harper_2025}, created using the destriper-only solution (left) and the gain-corrected solution described in this paper (right). The bright point source towards the upper end of the right plot is the QSO J0530+13, and the diffuse structure in the center of the right image is the B30 cloud.}
    \label{fig:galactic_maps}
\end{figure*}
\subsection{CMB removal}
Analysis of the COMAP CO-LIM data shows a slight residual of continuum emission from both point sources and the CMB when the linear filter is employed. This residual becomes stronger if masking of the system temperature spikes is only performed \textit{after} the linear filter is applied. This is illustrated in the left panel of Figure \ref{fig:CMB}, which shows a detector- and frequency-averaged COMAP map, with CMB contours. This effect is enhanced by the fact that channels with a positive continuum residual also happen to be in channels that are less noisy, as the system temperature of each channel decides not only its noise level, but also the sign of the continuum residual after the linear fit (see Fig.~\ref{fig:bad_weather}). Since we are averaging our entire $\SI{8}{GHz}$ band, the white noise is substantially suppressed, and we end up with a quite sharp image of the continuum residuals.

Since the low-level COMAP pipeline filters out the total offset in both angular and frequency dimensions, the resulting maps are almost exactly zero-centered, and averaging across frequencies does not yield an easily interpretable result. These plots are therefore mostly a curiosity that demonstrates that, when the naive linear fit is applied, there is indeed leakage of continuum into the final data. As the noise level in individual frequency channels is too high to perform a similar analysis in the full 3D maps, they nonetheless serve as a useful proof-of-concept for the method presented in this paper.

The middle panel of Fig.~\ref{fig:CMB} shows what happens if we apply system temperature spike masking before the linear fit, as done in the current COMAP CO-LIM analysis \citep{Lunde_2024}. In this case, the residual from both CMB and point sources drops significantly but remains present.

The last panel of Fig.~\ref{fig:CMB} shows the frequency-averaged map obtained using the technique described in this paper. There is no visually apparent correlation with the CMB map, and no visual residual of the point sources. The overall noise level is higher, but this just means that this method does not preserve the zero-centering of weighted co-added maps as well as the linear filter, which is not a concern, and does not indicate a higher noise level in the full 3D maps.

\section{Application to continuum science}

For the Galactic continuum science, $1/f$ gain fluctuations have been the dominant source of large-scale noise and uncertainty. The linear model of Eq.~\eqref{eqn:model_linear} cannot be applied, as this would also remove most of the Galactic continuum signal. Typically, when taking observations of diffuse continuum emission at microwave and radio frequencies $1/f$ gain fluctuations are controlled by using a reference load in a correlation receiver or switched receiver design. The COMAP pathfinder instrument, however, is a total power radiometer and was not optimized for diffuse continuum observations, so $1/f$ noise mitigation must be done during data processing. 

There are essentially two ways of employing our method to extract a cleaner continuum signal. The more aggressive approach is to directly treat $\delta T_t$ and $\delta T_t \alpha_t \bar \nu$ as a solution to the continuum signal, which in theory yields the highest signal-to-noise. The more conservative approach is to use the technique only to estimate and filter out the gain fluctuations, and treat $d_{\nu,t}^\mathrm{clean} = d_{\nu,t} - \delta G_t$ as the solution. The latter approach is safer and more likely to be unbiased, as it is only sensitive to leakage between $\delta G_t$ and $\delta T_t$, while the former approach is also sensitive to leakage between the residual and $\delta T_t$. The COMAP Galactic science pipeline, therefore, currently employs the latter, safer approach.

\subsection{COMAP Galactic Plane Survey results}

The first continuum results from COMAP, presented in \citet{Rennie_2022}, were of the inner Galactic plane between longitudes $20^\circ < \ell < 40^\circ$ and at latitudes of $|b| < 3^\circ$. These initial results were presented without the gain filter technique discussed in this paper. Instead, a destriper was employed during mapmaking. While this is a common method for $1/f$ mitigation, it accounts only for temporal noise correlations and does not exploit the spectral behavior of the signals, as the method we propose does.

For the bright inner Galactic regions, destriping mapmaking was sufficient. However, for much fainter regions, such as those of the dark-cloud B30 in the $\lambda$-Orionis ring \citep{Harper_2025}, destriping alone was found not to be enough to mitigate the $1/f$ noise. In Figure~\ref{fig:galactic_maps} we show the difference between the same set of observations of B30 presented by \citet{Harper_2025} at 27\,GHz when the gain filter is and is not applied. The right panel shows a visibly far cleaner map, and both the central B30 cloud and the point source in the top of the field are hard to discern in the destriper-only map.

\begin{figure}
    \centering
    \includegraphics[width=\linewidth]{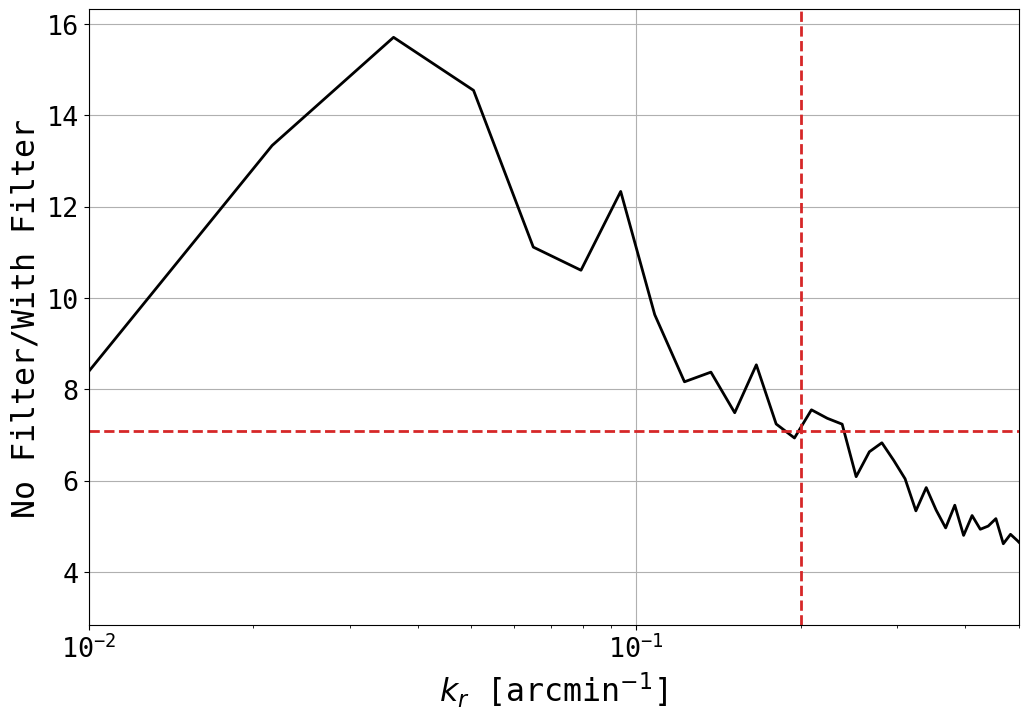}
    \caption{Figure showing the ratio of the map power spectra of a Galactic observation without the gain filter and with the gain filter. The angular scale of 5 arcminutes, approximately the size of the COMAP beam, is shown as red dashed lines.}
    \label{fig:GPS_noise}
\end{figure}

The ratio of the 2D radial power spectra of the noise in the two maps is shown in Figure~\ref{fig:GPS_noise}. We find that the noise power at the beam scale is approximately reduced by a factor of 7 when using the gain filter, and a factor of almost 16 at 30\,arcmin scales. In the map domain, this equates to an improvement factor in the noise of between 2.5 and 4 times, depending on the scale. 

\section{Conclusion}
We have presented an algorithm for jointly fitting the gain fluctuations and continuum signal of a total power spectrometer. The technique relies on an instrument with a system noise temperature that exhibits moderate spectral features that are temporally stable and accurately measured. We exploit these spectral features to build a three-parameter model that describes the gain and continuum contributions.

The COMAP Pathfinder instrument shows a series of sharp peaks in its system temperature. We have demonstrated that these spikes are temporally stable enough to act as calibrators for the gain, allowing the separation of gain noise and continuum signal. We expanded the model with a temporal prior on the $1/f$ gain fluctuations to reduce parameter degeneracies, thereby improving the fit's noise properties. The model was demonstrated on Jupiter scans, where the planet itself was almost solely contained within the continuum term, with less than $1\%$ leakage into the residual.

The technique can be applied in two different ways to COMAP science. For CO-LIM observations, the technique can be used to accurately remove \textit{both} gain and continuum fluctuations. This slightly improves upon the linear model currently employed by the CO-LIM pipeline, which does not fully account for continuum signals. We demonstrated this by more accurately subtracting CMB leakage in the CO-LIM maps.

The most effective application is to COMAP Galactic observations, where the technique was used to \textit{separate} gain fluctuations from a continuum signal. This allows for the subtraction of correlated noise with minimal bias to the continuum measurement. Simulations were used to show that even with a severely misestimated system temperature, the amount of leakage between the gain and continuum signals is limited to $\sim 5\%$. We also show that the typical misestimation of the COMAP system temperature is very likely lower than this.

Finally, the algorithm's effectiveness was demonstrated on Galactic science observations, significantly suppressing correlated noise. The produced maps had up to 4 times lower noise than those produced with only a destriper.

\begin{acknowledgements}
    We acknowledge support from the Research Council of Norway through grants 251328, 274990 and 357701.

    The current work has received funding from the European Union’s Horizon 2020 research and innovation programme under grant agreement numbers 819478 (ERC; \textsc{Cosmoglobe}), 101165647 (ERC; \textsc{Origins}), and 101141621 (ERC; \textsc{Commander}). This article reflects the views of the authors only. The funding body is not responsible for any use that may be made of the information contained therein. This research is also funded by the Research Council of Norway under grant agreement number 344934 (YRT; \textsc{CosmoglobeHD}).

    This material is based upon work supported by the National Science Foundation under Grant Nos.\ 1517108, 1517288, 1517598, 1518282, 1910999, and 2206834, as well as by the Keck Institute for Space Studies under ``The First Billion Years: A Technical Development Program for Spectral Line Observations''.

    CD and SEH acknowledge funding from a Science and Technology Funding Council (STFC) consolidated grant (ST/P000649/1) and a United Kingdom Space Agency (UKSA) grant (ST/Y005945/1) funding LiteBIRD foreground activities. GAH acknowledges the funding from the Dean’s Doctoral Scholarship by the University of Manchester.
\end{acknowledgements}

\bibliographystyle{abbrvnat}

\bibliography{references}

\clearpage
\appendix
\section{Maximum likelihood derivation}\label{app:ML_sol}
\subsection{Base model}

Equation~\eqref{eqn:final_model} can be written in matrix form as
\begin{equation}
    \vec{d} = \tens{P} \vec{a} + \vec{n}
\end{equation}
where $\mathbf{d}$ is the data flattened into a vector of size $N_t N_\nu$, $\mathbf{a}$ is a $3 N_t$ vector containing the solutions to the three parameters, $\mathbf{n}$ is the noise, and $P$ is a sparse $[N_t N_\nu,\, 3 N_t]$ matrix which maps $\mathbf a$ onto $\mathbf d$. We will assume that the noise $\mathbf n$ is uncorrelated and uniform, since both sources of correlated noise are part of our data model, and $\mathbf n$ then only accounts for any residual noise.

The matrix $P$ in practice looks like
\begin{equation}
    \tens{P} = 
    \begin{pmatrix}
        1  &  \bar{T}_0^{-1}  &  \bar{T}_0'^{-1} & 0 & 0 & 0 & 0 & \cdots \\
        0  &  0           &  0           & 1  &  \bar{T}_0^{-1}  &  \bar{T}_0'^{-1} & 0 & \cdots \\
        \vdots  & \vdots & \vdots & \vdots  & \vdots & \vdots & \vdots \\
        1  &  \bar{T}_1^{-1}  &  \bar{T}_1'^{-1} & 0 & 0 & 0 & 0 & \cdots \\
        0  &  0           &  0           & 1  &  \bar{T}_1^{-1}  &  \bar{T}_1'^{-1} & 0 & \cdots \\
        \vdots  & \vdots & \vdots & \vdots  & \vdots & \vdots & \vdots & \ddots \\
    \end{pmatrix}
\end{equation}
while $\mathbf a$ takes the form
\begin{equation}
    \vec{a} =
    \begin{pmatrix}
        \delta G_0 \\
        \delta T_0 \\
        \delta T_0'\\
        \delta G_1 \\
        \delta T_1 \\
        \delta T_1'
    \end{pmatrix},
\end{equation}
where we have, for compactness, defined $\bar{T}_\nu' = \bar{T}_\nu/\bar \nu$ and $\delta T_t' = \alpha_t \delta T_t$.
This defines a standard linear system, the log-likelihood of which takes the form

\begin{equation}\label{eqn:logL}
    \log \mathcal{L} \propto (\vec{d} - \tens{P}\vec{a})^T\tens{N}^{-1}(\vec{d} - \tens{P}\vec{a}).
\end{equation}
Minimizing Eq.~\ref{eqn:logL} with respect to $\mathbf a$ then yields

\begin{equation}
    \dv{\vec{a}} \log \mathcal{L} = 0 \quad \Rightarrow \quad \tens{P}^T \tens{N}^{-1} \tens{P} \vec{a} =  \tens{P}^T \tens{N}^{-1} \vec{d}.
\end{equation}

We assume a diagonal and constant noise covariance matrix $\tens{N} = \tens{I} \sigma^2$, where $\mathrm I$ is the identity matrix and $\sigma$ is the time-independent noise standard deviation. Note that, as we are explicitly considering the 1/f gain fluctuations as part of our data model, it does not enter the noise covariance term, as is common when considering correlated noise in a maximum likelihood problem. This makes $\tens{P}^T \tens{N}^{-1} \tens{P}$ a block-diagonal matrix, where only a series of $3\times3$ matrices needs to be inverted to produce the solution
\begin{equation}
    \vec{a} = (\tens{P}^T \tens{N}^{-1} \tens{P})^{-1} \tens{P}^T \tens{N}^{-1} \vec{d}
\end{equation}

\subsection{Base model + prior}
With the inclusion of a prior on $\mathbf a$ we get the log-posterior distribution 
\begin{equation}\label{eqn:logL_prior}
    \log \mathcal{P} \propto (\vec{d} - \tens{P}\vec{a})^T\tens{N}^{-1}(\vec{d} - \tens{P}\vec{a}) + \vec{a}^T  \tens{S}^{-1} \vec{a},
\end{equation}
where $S$ is the covariance of the prior on $\mathbf a$. This slightly changes the solution equation to become:
\begin{equation}
    \vec{a} = (\tens{P}^T \tens{N}^{-1} \tens{P} + \tens{S}^{-1})^{-1} \tens{P}^T \tens{N}^{-1} \vec{d},
\end{equation}
where $\tens{S}$ is a diagonal matrix in the Fourier domain, defining the power spectrum prior on $\vec{a}$. As noted in the main text, we place only a prior on $\delta G_t$, and not $\delta T_t$.

\section{Accuracy of $\bar T_\mathrm{\nu}$}\label{app:Tsys_all_uncertainties}
The accuracy of the method described in this paper relies on having an accurate measurement of the mean system temperature $\bar T_\nu$. This again relies on accurate and sufficiently frequent calibrations, which are used to determine $\bar T_\nu$. The COMAP telescope performs a calibration once every hour by rotating a vane of microwave-absorbing material across the entire field-of-view of the detectors. This calibration typically lasts for 5 to 10 seconds.

This appendix explores first the expected measurement uncertainty of $\bar T_\nu$, and then how well we expect this measured value to hold within the 1-hour window between calibrations.

\subsection{Measurement uncertainty}\label{app:Tsys_direct_uncertainty}
The mean system temperature of the COMAP Pathfinder instrument is calculated as \citep{Lamb_2022, Foss_2022}
\begin{equation}\label{eqn:app-Tsys}
    \bar T_\nu = \frac{T^\mathrm{amb}_\nu - T^\mathrm{CMB}_\nu}{P^\mathrm{amb}_\nu/P^{sky}_\nu - 1},
\end{equation}
where $T^\mathrm{amb}_\nu$ is the (ambient) temperature of the calibration vane, $T^\mathrm{CMB}_\nu$ is the brightness temperature of the CMB, $P^\mathrm{amb}_\nu$ is the averaged observed power during the calibration measurement, and $P^{sky}_\nu$ is the mean power observed during the scan being calibrated.

A $1\%$ uncertainty in the measured ambient temperature $T^\mathrm{amb}_\nu$ would lead to a $\approx 1\%$ uncertainty also in $\bar T_\nu$ (as $T^\mathrm{amb}_\nu \gg T^\mathrm{CMB}_\nu$). The ambient temperature is measured by a sensor on the telescope that logs two measurements per second, quantized to $\SI{0.2}{K}$. This introduces an uncertainty of up to $\SI{0.2}{K}$, or $\SI{0.2}{K}/\SI{300}{K}\approx 0.07\%$. 

This only accounts for measurement uncertainty. A mismatch between the sensor temperature and the emitted brightness temperature of the vane would lead to an additional uncertainty. Although design efforts were put in place to minimize any such mismatch \citep{Lamb_2022}, the total such discrepancy is unknown.

There will also be an uncertainty in the measurements of $P^\mathrm{amb}_\nu$ and $P^{sky}_\nu$, from the noise observed by the instrument. The uncertainty in the former term will dominate, as it is calibrated over a 5 second interval, compared to the 5 minutes of the latter. Propagating an uncertainty in $P^{amb}_\nu$ to $\bar T_\nu$ gives
\begin{equation}
    \sigma_{\bar T_\nu} = \qty|\pdv{\bar T_\nu}{P^{amb}_\nu}|\sigma_{P^\mathrm{amb}} = \frac{T^\mathrm{amb}}{\qty(P^\mathrm{amb}_\nu/P^{sky}_\nu - 1)^2}\frac{1}{P^{sky}_\nu}\sigma_{P^\mathrm{amb}}.
\end{equation}
The uncertainty of $P^\mathrm{amb}_\nu$ can be calculated as
\begin{equation}
    \sigma_P = \frac{P^\mathrm{amb}_\nu}{\sqrt{\Delta\nu \tau}} = \frac{P^\mathrm{amb}_\nu}{\sqrt{\SI{2}{MHz}\cdot \SI{5}{s}}} \approx 0.00032\cdot P^\mathrm{amb}_\nu.
\end{equation}
Using Eq.~\ref{eqn:app-Tsys} we get
\begin{equation}
    \frac{P^\mathrm{amb}_\nu}{P^{sky}_\nu} =\frac{T^\mathrm{amb}_\nu - T^\mathrm{CMB}_\nu}{\bar T_\nu} + 1 \approx \frac{\SI{300}{K} - \SI{2.7}{K}}{\SI{40}{K}} + 1 \approx 8.5,
\end{equation}
we get that
\begin{equation}
    \sigma_{\bar T_\nu} \approx \frac{\SI{300}{K}}{\qty(8.5 - 1)^2}\cdot 8.5 \cdot 0.00032 \approx \SI{0.015}{K}
\end{equation}
This corresponds to an uncertainty of $\SI{0.015}{K}/\SI{40}{K} \approx 0.04\%$.

In conclusion, the direct measurement accuracy of $\bar T_\nu$ is very high, expected to be at the $\sim 0.1\%$ level, with uncertainty contributions from both the measurement accuracy of the ambient load $P^\mathrm{amb}$, and the measurement accuracy of the ambient temperature $T^\mathrm{amb}$. This does not account for mismatches between the ambient temperature and the brightness temperature of the calibration vane, which are hard to quantify.

\subsection{Time stability}\label{app:Tsys_stability}
\begin{figure}
    \centering
    \includegraphics[width=\linewidth]{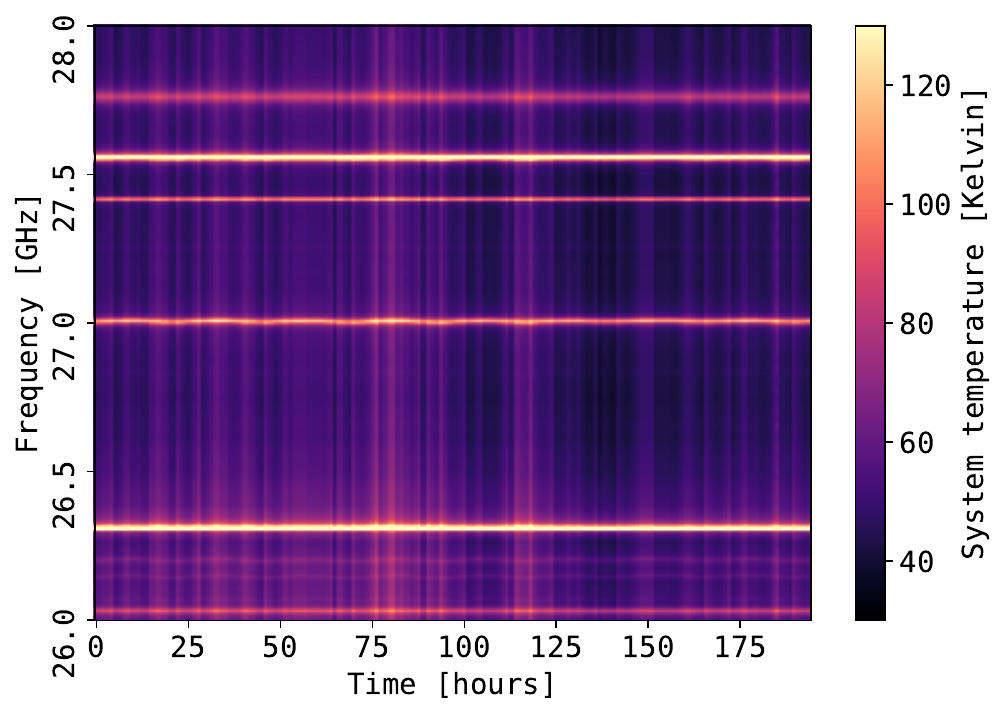}
    \caption{The mean system temperature $\bar T_\nu$, as measured by hourly calibrations. The figure contains 190 consecutive calibrations, or about a week of continuous observation. The system temperature spikes are very stable in time and do not appear to shift a lot between calibrations.}
    \label{fig:Tsys_over_time}
\end{figure}
Regardless of how precise the measurement of the mean system temperature is, our method can break if $\bar T_\nu$ changes meaningfully within the hour between calibrations. Note that a multiplicative change, of the form $c_t \cdot\bar T_\nu$, will not affect the shape of the templates. This change will be absorbed by the continuum parameter, $\delta \hat{T}_t \approx c_t \delta T_t$ while the gain parameter $\hat G_t$ remains unbiased. Whether this is a problem or not depends on how the technique is used.

The ability to constrain our model primarily comes from the integrated area under the system temperature spikes, which is the primary way in which the degeneracy between $\delta G_t$ and $\delta T_t$ is broken. There are then generally two ways in which a change in $\bar T_\nu$ can break our method: (i) If the system temperature spikes quickly drift in frequency between channels. (ii) If the system temperature spikes fluctuate meaningfully in amplitude, \textit{relative to} other channels.

Although we do not have a way of measuring the changes in $\bar T_\nu$ at timescales shorter than our calibration frequency, we can study how much it changes in between calibrations, as this is likely to provide an upper bound on how large changes are on smaller timescales.

Figure \ref{fig:Tsys_over_time} shows every hourly measurement of $\bar T_\nu$ over about a week of continuous observation. The most critical failure of our method would come from fast changes in the location of the system temperature spikes, as these are the primary source of constraining power. As we can see, the spikes do not tend to drift a lot over time. Only in 23\% of cases do the spikes change their peak channel between two hourly calibrations, and when they do, it is almost always by only a single channel. When a change in peak channel occurs between two subsequent calibrations, the average change is only 1.15 channels, out of the total 4096. Although what matters to our technique is how much the system temperature varies within a scanning period, and not at longer timescales, these findings show that the system temperature, and especially the spikes, are generally quite stable.

The figure also reveals some general fluctuations in the overall amplitude of the system temperature, due to physical temperature changes. As these changes affect all channels, they do not significantly alter the shape of the $\bar T_\nu^{-1}$ template, which is the important part for the purposes of our technique.

Finally, we assess how much the integral under the system temperature spikes drifts over time. Using the hourly calibrations, the integrated area of both the spiked channels and of the remaining ``flatter'' channels is calculated. The ratio of these integrals is then calculated and compared for all consecutive hourly calibrations. The mean absolute difference in this ratio between consecutive calibrations was then normalized by the average ratio over all calibrations, yielding the metric
\begin{equation}
    \frac{\left\langle \qty| A^\mathrm{spikes}_{i+1}/A^\mathrm{flat}_{i+1} - A^\mathrm{spikes}_{i}/A^\mathrm{flat}_i|\right\rangle}{\left\langle A^\mathrm{spikes}/A^\mathrm{flat}\right\rangle}.
\end{equation}
The average area difference between two calibrations was 2.1\% relative to the total area. This is significantly larger than the measurement uncertainty on $\bar T_\nu$. Note that changes on shorter timescales are likely smaller than between hourly calibrations, and this value should therefore be interpreted as an upper limit on the average drift in the system temperature spikes.

\begin{figure}
    \centering
    \includegraphics[width=0.99\linewidth]{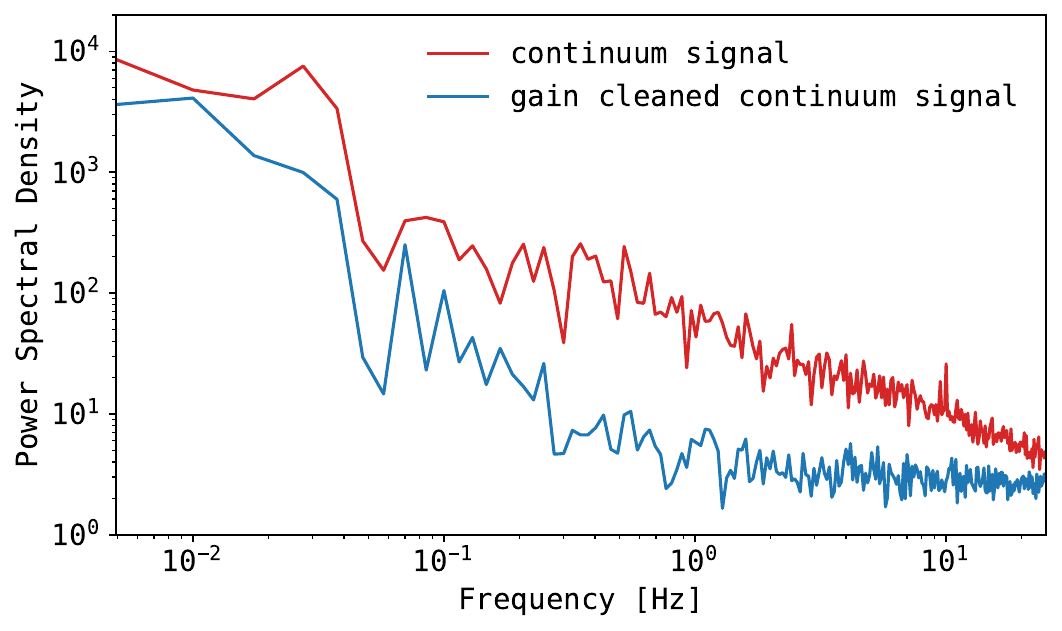}
    \caption{Continuum power spectrum from averaging the entire $\SI{8}{GHz}$ COMAP bandpass before (red) and after (blue) subtracting the gain from the data. The filter was applied on a single scan not observing continuum sources, and the only expected continuum emission is therefore the atmosphere, which can be seen as a residual at low frequencies.}
    \label{fig:continuum_cleaning_PS}
\end{figure}
\section{Gain and atmospheric power spectrum}
In addition to the correlated $1/f$ noise from the amplifier gain fluctuations, an additional correlated noise contribution from the atmosphere is commonly found in sub-mm astronomical observations. These atmospheric fluctuations also typically follow a $1/f$ temporal spectrum, but with a steeper slope than gain fluctuations. The atmosphere is therefore expected to be the dominant noise contribution at very long timescales.

Figure~\ref{fig:continuum_cleaning_PS} shows the power spectrum of a random COMAP scan averaged over the entire $\SI{8}{GHz}$ band, before and after the gain-cleaning procedure was applied to it. The gain-cleaning removes a lot of power on intermediate time-scales, but as we approach the lowest frequencies, the power of the cleaned signal starts rising again. This demonstrates the algorithm's capability of distinguishing the continuum source (atmosphere fluctuations) from the gain, and additionally illustrates how the atmosphere dominates the noise budget at the very longest timescales.

\end{document}